\documentstyle[12pt]{article}
\newcommand{\be}{\begin{eqnarray}}
\newcommand{\ee}{\end{eqnarray}}
\newcommand{\nn}{\nonumber}
\newcommand{\V}{{{\cal V}}}
\newcommand{\R}{{{\cal R}}}
\newcommand{\I}{{{\cal I}}}
\newcommand{\A}{{{\cal A}}}
\newcommand{\B}{{{\cal B}}}
\newcommand{\con}{\mbox{$\,$\rule{1ex}{0.4pt}\rule{0.4pt}{1ex}$\,$}}

\newcommand{\dwedge}{\dot{\wedge}}
\newcommand{\kla}[1]{\left( #1\right)}
\newcommand{\til}{\tilde{~}}
\input{psfig}
\begin{document}
%
%
% Titelblatt
%
\title{Clifford Algebraic Remark on the Mandelbrot Set of
Two--Component Number Systems}
\author{Bertfried~Fauser\\
        Institut f\"ur Theoretische Physik \\
        Universit\"at T\"ubingen\\
        Auf der Morgenstelle 14
\thanks{e-mail: ptifu01@commlink.zdv.uni-tuebingen.de}}
\date{Tue-prep-95-07-07\\
      arch-ive/9507133}
\maketitle
%
% Abstract
%
\begin{abstract}
\begin{sloppypar}
\noindent
We investigate with the help of Clifford algebraic methods the
Mandelbrot set over arbitrary two--component number systems. The
complex numbers are regarded as operator spinors in {\bf
D$\times$spin(2)} resp. {\bf spin(2)}. The thereby induced (pseudo)
normforms and traces are {\it not} the usual ones. A multi
quadratic set is obtained in the hyperbolic case contrary to
\cite{Met}. In the hyperbolic case a breakdown of this simple
dynamics takes place.
\end{sloppypar}
\end{abstract}

\noindent
{\bf PACS: 05.45.+b; 02.40.Dr; 02.60.Cb}
\newpage
\section*{I Introduction}

The quadratic Mandelbrot set is the set of non divergent points
in the iteration $z_{n+1}:=z_n+c$,~$z_0:=(0,0)$,~$\forall
c$, over the binary numbers, with the same norm as in the
complex case. Recently a convergence proof for the quadratic
Mandelbrot set was given \cite{Met}. This set was discussed
numerically in \cite{Sen}. The puzzling effect, that by changing
only one sign in the iteration formula results in a completely
different not even chaotic pictures, was expressed by the term
''perplex numbers``, which are nothing but binary numbers
\cite{Maj,Fje}. Where the disappearance of the chaotic behavior
was expressed by ''The mystery of the quadratic Mandelbrot set``
in \cite{Met}.

In this note, we want to show how the two cases can be
understood in a Clifford algebraic framework. This provide us
with several advantages over the usual picture.

First we introduce the complex numbers as operator spinors
acting on a unit reference vector in the Euclidean vector space
{\bf E(2)}. This is in analogy with Kustaanheimo, Stiefel
\cite{Kus}, Lounesto \cite{Lou-mech} and the inspiring chapter 8
''Spinor Mechanics`` of Hestenes' ''New Foundation of classical
Mechanics`` \cite{Hes-mech}.

In a second step we utilize Clifford algebras with arbitrary
not necessarily symmetric or antisymmetric bilinear forms. Such
algebras where discussed by Chevalley \cite{Che}, Riesz
\cite{Rie}, Oziewicz \cite{Ozi} and  Lounesto \cite{Lou-bili} in
a mathematical and geometrical manner.

This procedure is a general tool and was used in \cite{Fau} in
an entire other context. Here the geometrical point of view
allows a generalization of the concept of operator spinors to
those vector spaces, which are equipped with general non
degenerate bilinear forms.

The third step stems from the observation, that the Clifford
algebra is strongly connected with a quadratic algebra. This
structure provides us with several existence and uniqueness
theorems. Separability provides us with a unique (pseudo)
normform, as with a unique trace \cite{Hah}.

Surprisingly we are left with a norm different from the one used
in \cite{Met}. But here the (pseudo) normform is compatible with
the algebraic structure and the geometrical meaning. These
(pseudo) normforms and traces are therefore preferable.

Using these norms, we clearly see other images in numerical
experiments. The obtained pictures are in agreement with the
physical situation at hand. Several special cases will be
discussed. The Mandelbrot dynamic is shown to be incompatible
with this structure and a better one should be derived from
inhomogeneous {\bf D$\times$Ispin} or {\bf Ispin} groups. The
breakdown of the dynamics in the hyperbolic case may be
interpreted as a decay process or an absorption of the particle
considered.

In section II we introduce the operator spinor for {\bf
E(2)}\footnote{See Hestenes \cite{Hes-mech} for an enlargement
to 3--dimensional space and the spinor gauge formulation of the
Kepler motion.}. In section III we introduce Clifford algebras
with nonsymmetric bilinear forms, described explicitly in
\cite{Ozi}. In section IV we recall several facts from the
theory of quadratic algebras as proposed in \cite{Hah}. In
section V we discuss the Mandelbrot set over arbitrary
two--component number systems using the (pseudo) normforms
induced by algebraic and geometric considerations. The
conclusion summarizes our results and compares it to other work
done.

\section*{II Complex Numbers as Operator Spinors on {\bf E(2)}}

In this section we introduce the concept of algebraic spinors in
the sense of Hestenes \cite{Hes-mech,HesSob} and others
\cite{Che,Rie}. Therein a geometrical meaning is given to
the complex numbers. Indeed they are commonly identified with
coordinates, which is quite obscure. There is a long quest in
mathematics to give a geometrical meaning to the complex entities.
In spite of their natural occurring in algebraic geometry they
remained somehow mysterious \cite{Kle}. An attempt in this direction
seems very useful in the light of the ubiquitous appearance of
complex numbers and {\it complex coordinates} in physics.

We choose a standard orthonormal basis $e_i$,~$i\in (1,2)$ of
$CL(\mbox{\bf E(2)},\delta)$ with the properties (algebraproduct
by juxtaposition)
\be
e_i e_i = e_i^2=1&&\mbox{normalized} \nn \\
e_ie_j+e_je_i=0,&& i\not= j;\quad\mbox{orthogonality.}
\ee

The standard involution (conjugation) may be defined by
\be
J &:& CL \rightarrow CL \nn \\
J &:& J(ab):=J(a)J(b) \nn \\
J &:& J\vert_{{\bf K}\oplus{\bf E(2)}}:=id_{\bf K}-id_{\bf E(2)}
\nn \\
  &&  J^2=id_{CL}.
\ee
The reversion is the main antiautomorphism defined by (see
\cite{BudTra})
\be
\til &:& CL\rightarrow CL \nn \\
\til &:& (ab)\til :=b\til a\til \nn \\
\til &:& \til\vert_{{\bf K}\oplus{\bf E(2)}}:=id_{
{\bf K}\oplus{\bf E(2)}}\nn \\
&& (\til)\til=id_{CL},
\ee
where {\bf K} and {\bf E(2)} are the images of the field (ring)
and the vector space (module) in $CL$. Because there is a natural
injection we will not distinguish between this pictures.

An algebra basis is given by
\be
\{X_I \} &:=& \{1,e_2\wedge e_1 ,e_1,e_2\}\label{X}
\ee
with
\be
e_i\wedge e_j&:=&\frac{1}{2}(e_i e_j +J(e_j) e_i) \nn \\
e_i \con e_j&:=&<e_i\vert e_j>:=\delta_{ij}=\frac{1}{2}(e_1
e_2-J(e_2) e_1).\label{con}
\ee
We may use the symbol '$i$` to denote $e_2\wedge e_1$, because of
the properties
\be
i^2&=&(e_2\wedge e_1)(e_2\wedge e_1)=e_2 e_1 e_2
e_1=-e_2^2e_1^2=-1\nn \\{}%
[i,X_I]_{deg\vert X_I\vert}&=&0.
\ee
Where $deg\vert X_I\vert$ means the {\bf Z}$_2$--grade of the
homogeneous element $X_I$. For (odd) even elements we have the
(anti)commutator. This graduation is induced by $J$ and has the
structure
\be
CL&=CL_+\oplus CL_-,
\ee
where the even (+) part constitutes a subalgebra, whereas the
odd (-) part has only a $CL_+$ module structure. In the {\bf
E(2)} case, $CL_+$ is generated by $\{1,i\}$ and is itself a
Clifford algebra isomorphic to the field ${\bf C}$ of complex numbers..

Now we want to emphasize the operator character of complex
numbers. Therefore we calculate the left action of the element
'$i$` on the base vectors
\be
i e_1 &=&  e_2\nn \\
i e_2 &=& -e_1,
\ee
which is a counter clockwise rotation by $\pi/2$. If we choose
an arbitrary unit reference vector $e$, $e^2=1$, we may
write the elements in $CL_+$ as
\be
CL_+\ni z&:=&x+y i,\quad x,y \in {\bf K}.
\ee
So we have
\be
z=ze^2=(x e+ y ie)e=v e
\ee
with $v \in {\bf E(2)}$. From
\be
\{e, ie\}_+=eie+iee=-ie^2+ie^2=0
\ee
we have $ e \perp ie$. Thus $\{e,ie\}$ span {\bf E(2)}, and
we are able to rotate the coordinate system by an orthogonal
transformation to achieve
\be
e   =e_1\nn \\
i e =e_2.
\ee
The map $z \rightarrow ze$ is a bijective map from $CL_+
\rightarrow {\bf E(2)}$, because $e$ is invertible. On the other
hand we may look on the map $z : {\bf E(2)} \rightarrow {\bf
E(2)}$ with $za \rightarrow v$ for arbitrary $a \in$ {\bf E(2)}
and fixed $z\in CL_+$.

Only with the choice $e=e_1$ we are able to interpret the scalar
and bivector part of an operator spinor to be coordinates.

The modulus may be defined as follows
\be
\vert z \vert^2 &:=& zz\til =(x+iy)(x-iy)=x^2+y^2.
\ee

$CL_+$ is isomorphic to the field ${\bf C}$ and therefore
algebraically closed. Thus it is possible to find a root
\be
z&=&w^2= zz\til\frac{z}{zz\til}=\vert z\vert^2 u^2,\quad
uu\til=1,\quad \vert z\vert \in \mbox{{\bf K}}\sim\mbox{{\bf D.}}
\ee
We can reformulate the map as
\be
ze&=&w^2e=\vert z\vert^2 u^2=\vert z\vert^2 ueu\til.
\ee
This decomposes the left action of $z$ into a dilation and a
spinorial rotation. Because of $u\til=u^{-1}$, $u$ is a {\bf
spin(2)} transformation. The transformation obtained by the left
action of $u$ is of half angle type.

This point of view is independent of the special vector $e$ and
emphasizes as well the operator character of the iteration
formula. Indeed the iteration is a sequence of maps from ${\bf
E(2)} \rightarrow {\bf E(2)}$, where $a \in {\bf E(2)}$ is given
by $z_n(e)$.

This works well, because $CL_+ \cong {\bf C}$ is algebraically
closed and products as well as sums of $CL_+$ elements yields
new $CL_+$ elements, which can be interpreted as new operator
spinors $z^\prime$. In the general case, as in the hyperbolic
one, only the multiplicative structure forms a (Lipschitz) group
{\bf D}$\times${\bf spin(p,q)}, whereas the additive group is in
general incompatible with the geometric structure.
For example we find two timelike vectors in the forward
light cone, which become space like when added.

A physically sound dynamic model should therefore have an
invariant, i.e. multiplicative structure. By studying
multiplicative structures in vector spaces admitting one higher
dimension and performing a split \cite{HesSob,HesZie} one can
achive affine transformations with {\bf D}$\times${\bf
Ispin(p,q)}. For clarity and brevity, as well as for comparison
with the results of the literature we omit this complication.
However we take it into account, when we perform the actual
calculations.

\section*{III Clifford Algebras for Arbitrary Bilinear Forms}

In this section we give a brief account on Clifford algebras
with arbitrary bilinear forms. It's main purpose is to show the
connection to quadratic algebras. For a somehow polemic
discussion of Grassmann or Clifford algebra as a basic tool in
physics we refer to \cite{Ozi,Hes-clif}.

At first glance it is surprising to have non symmetric bilinear
forms in Clifford algebras, because in the usual approach
\cite{Dir} they arise naturally with symmetric bilinear or
sesquilinear forms. The situation looks even more puzzling when
noticing the universality of Clifford algebras, usually stated
as: ''There is up to isomorphisms only one unique algebraic
structure compatible with a bilinear form of signature (p,q)
on the space \V``.

Why is it worse to study isomorphic structures? In \cite{Fau} it
was demonstrated that the physical content of a theory depends
sensible on the embedding $\bigwedge\V\rightarrow CL(\V,B)$ of
the Grassmann or exterior algebra into the Clifford algebra.
For example normalordering is exactly such a change of this embedding.

Here we will give the connection between such changes and
the properties of quadratic algebras. They provide us with
a deeper geometric understanding as described in section IV.

Chevalley \cite{Che} was the first who decomposed the Clifford
algebra product into parts. These parts, then explicitly,
exhibit the twofold structure of the Clifford elements.

One part acts like a derivation on the space $\bigwedge\V$ of
exterior powers of $\V$. Especially if $\omega_x=x\con$ is a
form of degree $1$ it constitutes a form on $\V$ into the field
{\bf K} ($x,y \in \V$,~$\con : \V\rightarrow \V^\ast$),
\be
\omega_x(y)=x\con y&:=&B(x,y) \in {\bf K}.
\ee
Thus $\con$ is a dualisomorphism parameterized by $B$.
The second part of the Clifford product is simply the exterior
multiplication.

Marcel Riesz \cite{Rie} reexpressed the
contraction and the exterior multiplication with a grade
involution $J$ and the Clifford product. He obtained for char
{\bf K} $\not=$ 2, $x\in\V,\, u\in CL$
\be
x\con   u&:=&\frac{1}{2}(xu-J(u)x)\nn \\
x\wedge u&:=&\frac{1}{2}(xu+J(u)x).
\ee
One is able to extend these operations to higher degrees of
multivectors by $(a,b,c \in \V,\, X\in CL)$
\be
(a\wedge b)\con X&=&a\con(b\con X)\nn \\
a\con(bc)&=&(a\con b)c+J(b)(a\con c).
\ee
Associativity of the wedge product and other properties are
shown in \cite{Rie}. A detailed account on such properties is
given in \cite{Ozi}.

In the following it is convenient to use explicit not
necessarily normed or orthogonal generating sets. Especially for
low dimensional examples this will be useful. Therefore we
define the left--contraction and the exterior product
as ($\V=span\{e_1,\ldots,e_n\}$)
\be
e_i\con e_j&:=&B(e_i,e_j)\cong [B_{ij}]\nn \\
e_i\wedge e_j&:=&\frac{1}{2}(e_i e_j- e_j e_i)
\ee
with the standard grade involution $J(e_i)=-e_i$.

Next we may decompose $B$ into symmetric and antisymmetric part
\be
G+F&:=&B \nn \\
G^T& =&G \nn \\
F^T&=&-F
\ee
where ${}^T$ is the matrix transposition. In the case of an
algebra over the complex numbers one has to use hermitean
adjunction, but here we deal exclusively with real algebras.

We specialize now to 2 dimensions. Then we are left with 4
parameters, 3 of the symmetric and 1 of the antisymmetric part.
\be
[G]&=&\kla{\begin{array}{cc}
            G_{11} & G_{12} \\
            G_{12} & G_{22}
           \end{array}} \nn \\
{}[F]&=&\kla{\begin{array}{cc}
            0      & F_{12} \\
            -F_{12}& 0
           \end{array}}.
\ee
The even part of this algebra satisfies a quadratic equation for
every element $z=x+y e_2\wedge e_1$ where
\be
(e_2\wedge e_1)^2&=&(e_2 e_1-B_{21})(e_2\wedge e_1) \nn \\
                 &=&e_2(e_1\con (e_2\wedge e_1))-B_{21}e_2\wedge
                 e_1 \nn \\
                 &=&e_2(B_{12}e_1-B_{11}e_2)-B_{21}e_2\wedge
                 e_1 \nn \\
                 &=&(B_{12}-B_{21})e_2\wedge
                 e_1-(B_{11}B_{22}+B_{12}B_{21}) \nn \\
                 &=&-2F_{21}e_2\wedge e_1-det(B). \label{q1}
\ee
Now we expand $det(B)$
\be
det(B)&=&B_{11}B_{22}-B_{12}B_{21} \nn \\
      &=&G_{11}G_{22}-(G_{12}+F_{12})(G_{12}-F_{12}) \nn \\
      &=&F_{12}^2+det(G) \label{det}
\ee
and we arrive at
\be
(e_2\wedge e_1)^2&=&2F_{12}e_2\wedge e_1-det(G)-F_{12}^2.
\ee
If there is no antisymmetric part (e.g. $F_{12}\equiv 0$), we
specialize to
\be
(e_2\wedge e_1)^2&=&-det(G)
\ee
and the determinant of $G$ describes the geometry at hand.
If $det(G)$ is positive we arrive at a (anti) Euclidean geometry.
In the negative case the geometry is hyperbolic.

We are free to choose an other algebra basis via a new doted
wedge product defined by
\be
e_i\dwedge e_j&:=&F_{ij}+e_i\wedge e_j.
\ee
Therefore we have incorporated the whole antisymmetric part in
the wedge product and the Clifford product decomposes as
\be
e_i e_j&=&B_{ij}+e_i\wedge e_j \nn \\
       &=&G_{ij}+e_i\dwedge e_j.
\ee
We define a new set of elements spanning the algebra, which
reads in two dimensions
\be
\{ Y_J\} &:=&\{1,e_2\dwedge e_1,e_1,e_2\}\label{Y}
\ee
which leads to another quadratic relation
\be
(e_2\dwedge e_1)^2&=&(e_2\wedge e_1+F_{21})^2 \nn \\
       &=&(e_2\wedge e_1)^2+2F_{21}e_2\wedge e_1+F_{21}^2 \nn \\
       &=&2(F_{21}+F_{12})e_2\wedge e_1
       -det(G)-F_{12}^2+F_{21}^2 \nn \\
       &=&-det(G). \label{q2}
\ee
In this construction the linear term has been absorbed in the
doted wedge. The quadratic relation has simplified to the
homogeneous case discussed above, but now not as a special case.

Next we introduce a (pseudo) norm function on the
quadratic subalgebra. In the Clifford algebra there are several
such constructions, known as spinor norms \cite{Lou-spin,Cru}.

For vector elements $a\in \V$ every of the three following maps
has the image in the field {\bf K}. With the standard involution
$J(a)=-a$ we have
\be
a^2\rightarrow {\bf K};\quad
aJ(a\til)=aJ(a)\til=-a^2\rightarrow {\bf K};\quad
aa\til=a^2\rightarrow {\bf K}
\ee
But applying this to bivector elements, the first relation maps
not in the field, but is exactly the quadratic relation derived
above. If we allow arbitrary involutions, $a$ and $J(a)$ need
not be parallel vectors, as is the same for bivector elements.
In general we may not expect this map to be scalar valued. To
analyze the third map, we recognize
\be
(e_i\wedge e_j)\til &=& (e_i e_j-B_{ij})\til=e_j e_i-B_{ij} \nn \\
      &=&e_j\wedge e_i+B_{ji}-B_{ij}  \nn \\
      &=&e_j\wedge e_i-2F_{ij} \nn \\
      &=&- e_i\wedge e_j-2F_{ij}.
\ee
Thus we have with (\ref{q1})
\be
(e_i\wedge e_j)(e_i\wedge e_j)\til&=&-(e_i\wedge
   e_j)^2-2F_{ij}e_i\wedge e_j \nn \\
   &=&det(B)+2F_{ij}e_i\wedge e_j-2F_{ij}e_i\wedge ej \nn \\
   &=&det(B)
\ee
as with (\ref{q2}) in the same manner for the doted case
\be
(e_i\dwedge e_j)(e_i\dwedge e_j)\til &=&-(e_i\dwedge
e_j)(e_i\dwedge e_j)\nn \\
&=&det(G).
\ee
This reduces in the case of two dimensions to
\be
(e_2\dwedge e_1)(e_2\dwedge e_1)\til &=&det(G).
\ee
which is the discriminant of the quadratic equation (\ref{q2}).
If the discriminant is positive, there are two roots in the
algebra (over {\bf R}), whereas otherwise the field has to be
algebraically closed (i.e. {\bf C}), or extracting the root is
not possible.

We introduce the abbreviation $X \cong\{e_2\wedge e_1$ or
$e_2\dwedge e_1\}$, $a\cong \{2F_{12}$ or $0 \}$, $b\cong \{
det(B)$ or $det(G) \}$ and may
write the quadratic algebra as
\be
\frac{K[X]}{X^2-aX-b},
\ee
the polynom algebra generated by $X$ over the field ${\bf K}$
modulo the quadratic relation. This form is important for
comparison with the theory of quadratic algebras, but our aim
is the exposition of the connection to the geometric relations
encoded in this formula.

\section*{IV Quadratic Algebras, Conjugation and Special Elements}

In this section we give some results exposed in \cite{Hah},
which provide us with existence and uniqueness theorems. This
supplies more fondness to our somehow loosely construction above.

The constructions are valid in much more general settings, as
Clifford algebras over finite fields or over modules, which is
yet not needed but quite interesting.

Let $\R$ be a commutative ring and $\I$ an ideal of $\R$, then
$\R\rightarrow \R/\I=\A$ is in a natural way a $\R$--algebra.

A free quadratic algebra is obtained if one factorizes the
polynom algebra in the indeterminate $X$ over $\R$ by the ideal
$\I=(X^2-aX-b)$
\be
S&=&\frac{\R[X]}{X^2-aX-b}.
\ee
The identity is
\be
1_S&=&1+\I=1+(X^2-aX-b)
\ee
and because $r\in\R$, $r\rightarrow r1_S$ is injective we have
$\R\subseteq S$. A basis of $S$ as $\R$--module is given by
\be
\{1&,& v=X+\I=X+(X^2-aX-b) \}.
\ee
It follows that
\be
v^2&=&av+b.
\ee
We have several isomorphisms:
\be
S_1&=&\frac{\R[X]}{X^2-X}\cong \R\oplus \R
\ee
with the diagonal product map. $\phi : \R\oplus \R \rightarrow
S_1$ is an isomorphism and we denote $S_1$ as trivial quadratic
$\R$--algebra. This is the ''perplex`` case from above!
\be
S_2&=&\frac{\R[X]}{X^2}
\ee
constitutes the algebra of dual numbers.

Denoting the units of $\R$ as $\R^\ast$, we can formulate the
isomorphism criterion \cite{Hah}(1.1).

{\it Let $\R[X]/(X^2-aX-b)$ and $\R[X]/(X^2-cX-d)$ be quadratic
algebras over $\R$. Then
\be
\frac{\R[X]}{X^2-aX-b}&\cong&\frac{\R[X]}{X^2-cX-d} \nn
\ee
iff there exist elements $r\in\R$ and $u\in\R^\ast$ such that
(i) $ c=ua+2r$ (ii) $d=u^2b-rua-r^2$.}

Now, has $X^2-aX-b$ a root, say $\gamma$ in $\R$ then we have
\be
X^2-aX-b&=&(X-\gamma)(X-(a-\gamma)) \label{root}
\ee
and $a-\gamma$ is another root (in $\R$) of the quadratic
equation. If $a-\gamma=\gamma$ then $\gamma$ is a double root.
We can state the following
\be
\begin{array}{ccccl}
(1) & \frac{\R[X]}{X^2-aX-b} &\cong& \frac{\R[X]}{X^2-taX-t^2b} &
\mbox{for $t\in \R^\ast$} \\
(2) & \frac{\R[X]}{X^2-aX-b} &\cong& \frac{\R[X]}{X^2-cX} &
\Leftrightarrow X^2-aX-b\quad \mbox{has a root in $\R$} \\
(3) & \frac{\R[X]}{X^2-aX-b} &\cong& \frac{\R[X]}{X^2} &
\Leftrightarrow X^2-aX-b\quad \mbox{has a double root in $\R$}
\end{array} \nn
\ee
{\bf Examples:}
\begin{enumerate}
\item $\R\cong {\bf C}$: The only quadratic algebras are ${\bf
C}[X]/(X^2-X)$ the complex trivial algebra and ${\bf
C}[X]/(X^2)$ the algebra of complex dual numbers, because of the
existence of roots for every  $X\in{\bf C}$.

\item $\R\cong {\bf R}$: We have three cases, because negative
numbers posses no roots in ${\bf R}$.
\be
S &\cong&
\left\{\begin{array}{ccl}
\frac{{\bf R}[X]}{X^2-X} & a^2+4b > 0 & \mbox{trivial, ''perplex``} \\
\frac{{\bf R}[X]}{X^2}   & a^2+4b = 0 & \mbox{dual numbers} \\
\frac{{\bf R}[X]}{X^2+1}\cong {\bf C} & a^2+4b < 0 &
\mbox{complex numbers}
\end{array}\right.
\ee
Of course, $a^2+4b$ is the discriminant of the quadratic relation.

\item ${\R\cong {\bf Z}}$: Results in the infinitely many
isomorphism classes
\be
\frac{{\bf Z}[X]}{X^2-aX-b}&\cong&\left\{
\begin{array}{cl}
\frac{{\bf Z}[X]}{X^2-n} & \mbox{if $a$ is even} \\
\frac{{\bf Z}[X]}{X^2-X-n} & \mbox{if $a$ is odd}
\end{array}\right.
\ee
\end{enumerate}
It turns out, that simpler rings (as also finite Galois
fields) bear much more structure.

If $\alpha$ is an (anti) automorphism and $\alpha^2=id_\A$, then
$\alpha$ is an involution. Algebra homomorphisms which preserve
such a structure are called graded homomorphisms. In physics the
super symmetric transformations (mixing of Grassmann parity)
are not grade perserving and thus minor symmetric.

In a quadratic algebra we may introduce the involution $\sigma :
S \rightarrow S$ on the base
\be
\{ 1, v&=&X+(X^2-aX-b) \} \nn \\
1^\sigma &=& 1 \nn \\
v^\sigma &=& (a-v)
\ee
which results in
\be
(x+yv)^\sigma &=& (x+ya)+yv \nn \\
(x+yv)^{\sigma\sigma} &=& (x+yv).
\ee
This involution interchanges the roots of the quadratic relation
(\ref{root}). In the complex case this is the ordinary complex
conjugation.

Now this kind of involution is ''standard`` in
quadratic algebras and induces the {\bf Z}$_2$--grading of the
Clifford algebra.
Let $\A_i$, $i\in (0,1)$ be $\R$--submodules and $\A=A_0\oplus
\A_1$. As $\R1 \subseteq \A_0$, $\A$ is a {\bf Z}$_2$--graded
algebra. Elements $a\in \A_i$ ($\partial a=i$, grade of $a$) are
called homogeneous.

We have two possibilities to introduce tensor products in graded
algebras via
\be
\A {\otimes}_\R \B &:& (a\otimes b)(a^\prime \otimes
b^\prime):=(aa^\prime\otimes bb^\prime) \nn \\
\A {\hat\otimes}_\R \B &:& (a\otimes b)(a^\prime \otimes
b^\prime):=(-)^{\partial a^\prime\partial b}(aa^\prime\otimes
bb^\prime).
\ee
To the algebra $\A$ we find the opposite algebra $\A^{op}$ by
reversing the product, $(ab)^{op}=b^{op}a^{op}$, which is a map
from $\A$ into $\A^{op}$. Hence we construct the enveloping
algebra $\A^e:=\A\otimes_\R \A^{op}$, as a ($\A,\A$)--bimodule.
There is a unique homomorphism $\Phi : \A^e\rightarrow \A$,
which satisfies $\Phi ( a\otimes b^{op})=ab$. If there exists also
a homomorphism $\Theta : \A\rightarrow \A^e$ (coproduct) such that
$\Phi\Theta=id_\A$ then the algebra is called separable. It follows
then $\A^e=\A\otimes_\R\A^{op}=ker(\Phi)\oplus\Theta(\A)$. With
\cite{Hah}(2.1)(2.3) we state:
\begin{enumerate}
\item {\it $\A$ is a separable $\R$--algebra iff $\A$ has a
separability idempotent, $e=\Theta(1)$.}
\item {\it The separability idempotent of a free quadratic
algebra $S$ over $\R$ is unique.}
\end{enumerate}
Now it is possible to classify separable free quadratic
$\R$--algebras by introducing the group $Qu_f(\R)$. Therefore define
\be
Q&:=&\{(a,b) \vert a^2+4b\in \R^\ast\}
\ee
with the product
\be
(a,b)*(c,d)&:=&(ac,a^2d+c^2b+4bd)
\ee
and the quotients $[a,b]=(a, b)/\R^{\ast 2}$ form the group
\be
Qu_f(\R)&:=&\{[a,b]\vert (a,b)\in Q\}.
\ee
The cardinality of $Qu_f(\R)$ is the number of isomorphism
classes of (nontrivial) quadratic algebras, e.g.
\be
Qu_f({\bf C})&\cong&\frac{{\bf C}^\ast}{{\bf C}^{\ast^2}}=1 \nn \\
Qu_f({\bf R})&\cong&\frac{{\bf R}^\ast}{{\bf R}^{\ast^2}}={\bf Z_2}.
\ee
This can be extended to a graded version $QU_f(\R)$.

One observes a connection between grading and standard
involution via the special elements. To see this,
let $M=(\V,B)$ be the pair of a vector space with a bilinear
form (or a quadratic module), then we can build the Clifford
algebra $CL(M)=CL(\V,B)$. We get now \cite{Hah}(5.4)

{\it The decomposition $CL(M)=CL_0(M)\oplus CL_1(M)$ is a ({\bf
Z}$_2$) grading of $CL(M)$. $CL_0(M)$ is a subalgebra and
$CL_1(M)$ is a $CL_0(M)$ module.}

We call $\sigma$ a standard involution, if $\sigma$ is a
antiinvolution and $aa^\sigma\in\R$ $\forall a\in CL(M)$. In
this case we define a (pseudo) norm and a trace as
\be
nr(a)&:=&aa^\sigma (\in \R) \nn \\
tr(a)&:=&a+a^\sigma (\in \R). \label{norm}
\ee

An element $z\in CL(M)$ is a special element if $\{1,z\}$ is a
basis of the centralizer $Cen_{CL(M)} CL_0(M)=\{c\in CL(M) \vert
cd=dc\,\, \forall d\in CL_0(M)\}$. One can conclude
that\footnote{See \cite{Hah} (8.3)}
\begin{enumerate}
\item if rank $M$ is odd, then $z\in CL_1(M)$, $z^\sigma=-z$ and
$z^2=b$ with $b\in\R$.
\item if rank $M$ is even, then $z\in CL_0(M)$, $z^\sigma=a-z$
and $z^2=az-b$ with $a,b\in\R$, $a^2+4b\in\R^\ast$.
\end{enumerate}

{\it If $\gamma$ is a root of $X^2-aX-b$ in $S$, then $S$ has
the grading $S=S_0\otimes S_1$, where
\be
S_0&=&Cen_S(\gamma)=\{s\in S \vert \gamma s= s\gamma \} \nn \\
S_1&=&\{s\in S \vert \gamma s+ s\gamma=as \}.
\ee}

The grading is trivial if $\gamma$ is in the center of $S$.

These properties are strongly interwoven and can be used in
constructing representations of Clifford algebras \cite{Dim}.

The existence of special elements in a general setting is proved
in chapter 10 of \cite{Hah}. All this constructions are possible
in higher dimensions, but our above naive geometric
interpretation has then to be refined.

As a last topic we have a look at the representations of
Clifford algebras. A homomorphism $\Phi$ of $\R$--algebras
\be
\Phi &:& CL(M)\rightarrow End_S(P)
\ee
where $M$ is a quadratic module, $S$ a $\R$--algebra and $P$ a
right $S$--module is called an ($S$--)representation of $CL(M)$.
In \cite{Hah}(8.7)(8.8) the connection between division
algebras, gradings and the quadratic algebra is shown and
connected with the existence of roots in $S$. The quadratic
algebra $S$ appears as tensor factor in such representations.

Despite the universality of Clifford algebras, we need for
physical applications norms and traces, as the grading
(vector space and {\bf Z}$_2$). Therefore we have to distinguish
between such homomorphisms preserving this additional structures
and those doing this not. The free quadratic groups etc.
characterize the isomorphism classes available, where the kernel
of the factorization parameterizes distinct but isomorphic
representatives. Such a parameterization can be done equally well
by parameterizing the isomorphic ideals in constructing Clifford
algebras. This was done in \cite{Ozi-hecke}.

\section*{V The 2--dim. Case and Numeric Experiments}

In this section we discuss the 2--dim. case
over {\bf R}. Because of the even dimension we expect to have
inhomogeneous terms and a rich structure. Using the natural
given standard involution, we can construct (pseudo) norms and
traces as above and investigate the Mandelbrot set in each of the
three cases. We do not get a quadratic set, but a ''light cone``
structure in the hyperbolic case.

Let us choose a generating set $\{e_1, e_2\}$ as in (\ref{X}) or
in (\ref{Y}), and a bilinear form $B=G+F$ as above. We denote
the bivector element as $\gamma$. Here one has to take care,
because a change in the quadratic relation of $\gamma$ is
related with a change of the representation of the
algebra. In the same time this results in a redefinition of the
wedge, as above done by using the two extreme cases $\wedge$ and
$\dwedge$.

We interpret the iteration formula in the operational sense
explained in section II and plot all figures in the
$\{1,\gamma\}$--plane. A translation into $\V$ may be done as
explained also in section II. For comparison to other results
this is here {\it not} done.

With Sylvester's theorem we could achieve a diagonal form for the
symmetric part of $G$ with $G\equiv diag\{\pm 1, \pm 1\}$. But a
rescaling of the base vectors by $\sqrt{\vert G_{11}\vert}$ and
$\sqrt{\vert G_{22}\vert}$ would affect the magnitude of the
nonsymmetric part also. This is contained in the isomorphism
criterion and will therefore be done only in the quadratic
algebra and not in the whole Clifford algebra.

Define $S=\R[\gamma]/(\gamma^2-a\gamma-b)$. We get from (\ref{det})
\be
\gamma^2&=&2F_{12}\gamma-det(B) \nn \\
        &=&2F_{12}\gamma-det(G)-F_{12}^2.
\ee
Hence we set
\be
a&:=&2F_{12} \nn \\
b&:=&-det(B)=-det(G)-F_{12}^2.
\ee
The discriminant is connected to the metric via
\be
d\, =\, a^2+4b&=&-4det(G).
\ee
The discriminant of the quadratic relation is thus $-4$ times
the determinant of the symmetric part of the bilinear form!

If $det(G)=0$ we have 2 roots in the algebra $S$. If the algebra
is not dual (double root) we have $det(G)\not= 0$.

As explained in section IV, there exists a standard involution
if the algebra $S$ is separable. In our case this is the
reversion in $CL$. It is constructed in the base $\{1,\gamma\}$ by
\be
1^\sigma &=& 1 \nn \\
\gamma^\sigma &=& (a-\gamma)=2F_{12}-\gamma.
\ee
The inhomogeneous additive term is quite uncommon in usual
approaches to Clifford algebras. Let us emphasize, that in the
quantum mechanical case, where the field is complex, we are able
to find {\it always} an algebra isomorphism to achieve
$\gamma^{\prime\sigma}=\gamma^\prime$. Because there is only one
nontrivial isomorphism class. This may be an argument for using
complex numbers in quantum mechanics.

The iteration formula is now obtained as follows
\be
z&:=&z_x+z_y\gamma;\quad c:=c_x+c_y\gamma
\ee
and with
\be
z_{n+1}&:=&z_n^2+c
\ee
we obtain
\be
z_{x(n+1)} &=& z_{x(n)}^2+bz_{y(n)}^2+c_x \nn \\
z_{y(n+1)} &=& 2z_{x(n)}z_{y(n)}+az_{y(n)}^2+c_y.
\ee
The spinor (pseudo) norm is then given by
\be
nr^2(z)&=&zz\til=z_x^2-bz_y^2+az_xz_y.
\ee
We have:
\begin{itemize}
\item[i)] {\it
$nr^2(z)$ is positive definite in the complex isomorphism class}
\item[ii)] {\it
$nr^2(z)$ is positive semidefinite in the dual isomorphism class}
\item[iii)] {\it
$nr^2(z)$ is indefinite in the trivial isomorphism class
(hyperbolic case)}
\end{itemize}
{\bf Proof:} We distinguish two cases
\begin{itemize}
\item[a)] Suppose that $y\equiv 0$. We are left with
$nr^2(z)=x^2$, which is positive definite in $x$ and $nr^2$ is
semidefinite in case iii), by isomorphy to the case $a=b=0$.
\item[b)] Suppose $y\not= 0$. We introduce $\rho=x/y$ and have
\be
\frac{nr^2}{y^2}&=&\rho^2-a\rho-b\nn \\
                &=&(\rho-\frac{a}{2})^2-(\frac{a^2}{4}+b)\nn \\
                &\geq&-(\frac{a^2}{4}+b)=-d=4\, det(G).
\ee
Thus if the discriminant is negative the norm is positive
definite. If $d=0$, the norm may be zero for non--null
elements, that is positive semidefinite. Is $d > 0$ we are in the
indefinite hyperbolic case. \rule{1ex}{1ex}
\end{itemize}

The variable $\rho$ is connected via an arctan or arctanh to the
phase of $z$. But a polar decomposition is in the hyperbolic
case not obvious. See \cite{Fje} and notice the appearance of
the Klein group.

The line $\R 1_S$ is stabilized by $\sigma$ via $z^\sigma=z$
$\Rightarrow$ $y=0$. Whereas the trace maps $z$ onto $\R 1_S$ as
\be
tr(z)&=&\frac{1}{2}(z+z^\sigma)=x+\frac{a}{2}y.
\ee
We define the operator spinor Mandelbrot set as
\be
M&:=&\{c\vert z_0=(0,0),\quad nr^2(z_n^c) > 0\, \forall n,\quad
\lim_{n \rightarrow\infty}\, nr^2(z^c_n)\not\rightarrow\infty.\}
\ee

In the complex case we set the parameters as
\be
a&=&0 \nn \\
b &=& -1,\quad \rightarrow\quad \gamma^2=-1.
\ee
Hence
\be
d\, =\,a^2+4b&=&-4=-4\, det(G)\nn \\
det(G)&=&1,
\ee
which results in a (anti) Euclidean geometry. Because of the
positive discriminant we have always two roots in $S$. The norm
yields
\be
nr^2_{\bf C}(z)&=&x^2+y^2.
\ee
We are left with the ordinary Mandelbrot set as shown in figure
1. All pictures with $d= -4.0$, and arbitrary $a$ are isomorphic
without a rescaling. But the coordinate interpretation is no
longer possible. If one looks at $d = -4.0$, $a=1.0$ (figure
2.), one obtains up to an {\bf SO(2)} transformation the
bilinear form
\be
B&=&G+F\cong
\left(
\begin{array}{cc}
1 & \frac{1}{2} \\
0 & 1
\end{array}
\right).
\ee
Thus we have now
\be
\gamma e_1 &=& (e_2 e_1-B_{21})e_1=e_1^2 e_2=e_2 \nn \\
\gamma e_2 &=& \frac{1}{2}e_2-1 e_1.
\ee
The new transformation obtained by $\gamma$ : {\bf
E(2)}$\rightarrow$ {\bf E(2)} does not preserve angles, but areas.
The real axis ($\sigma$ invariant points) is not affected by this.
The result is the deformed set of figure 2.

If the parameter $d$ is changed, this results in a scaling of
the $\gamma$--axis. As in the above case the transition into the
vector space is not unique. If one chooses the 1$_S$--axis to map
onto $e_1$ then we arrive at a bilinear form like
\be
[B]&=&
\left(
\begin{array}{cc}
d & \frac{a}{2} \\
0 & 1
\end{array}
\right)
=\left(
\begin{array}{cc}
G_{11} & F_{12} \\
0 & 1
\end{array}
\right),
\ee
which is a special case of the parameterization
\be
[B] &=&d
\left(
\begin{array}{cc}
\lambda & \frac{a}{2} \\
0 & \frac{1}{\lambda}
\end{array}
\right).
\ee
This case is also isomorphic to the complex one, but this time
with an additional rescaling (figure 3). Areas are no longer
preserved.

The second case is the dual one. Hence the algebra $S$ is no
longer separable and degenerated to a 1--dim. scheme. The
corresponding parameters are
\be
a&=&0 \nn \\
b&=&0,\quad\rightarrow\quad\gamma^2=0.
\ee
The discriminant vanishes. This case ($X^2=0$ as ideal) results
in a degeneration of the dynamics. It is the limiting case of
the two other ones. The semidefinite norm is $nr^2(z)=(x-a/2\,
y)^2$, which is sensitive only to one direction. In the case
with $a = 0$ this is the $x$--axis, see figure 4. To form a
definite norm, and thus a physically meaning full situation, one
has to factor out the superfluous direction. So this case is
essential one dimensional. This can take place even if $B$ is
nondegenerate, but $G$ is still.
\be
[B]&=&
\left(
\begin{array}{cc}
G_{11} & F_{12} \\
-F_{12}& 0
\end{array}
\right)=
\left(
\begin{array}{cc}
G_{11} & 0 \\
0      & 0
\end{array}
\right)
+\left(
\begin{array}{cc}
0       & F_{12}\\
-F_{12} & 0
\end{array}
\right)
\ee
In a physical context this case should be called trivial, but
this has not to be confused with the classification of the
quadratic algebras, where this case is the dual one.

The hyperbolic (or ''perplex``) case is obtained with the
parameter setting
\be
a&=&0 \nn \\
b&=&1,\quad \rightarrow\quad \gamma^2=1.
\ee
The discriminant becomes
\be
det(G)=-1,
\ee
which corresponds to the hyperbolic geometry. Not every element
in $S$ has a root in $S$ and especially $\gamma$ has not.

In \cite{Met} the convergence was proved with the norm $nr_{\bf
C}$ from above. With our considerations we get contrary
\be
nr^2_{{\bf R}\oplus{\bf R}}&=&x^2-y^2.
\ee
We recover the ''light cone``, which one is used to find in
hyperbolic geometry. Hence $x$ is the timelike coordinate and
$y$ the space like. Backward and forward light cone enclose the
invariant real line ${\bf R} 1_S$.

As exposed in the introduction, the dynamic (iteration process)
does not respect this structure. So timelike elements will
become space like and vice versa. The pictures were done in such a
way, that the iteration halted immediately whenever an element got
space like.

The most surprising effect is, that the light cones become
separated by the multi quadratic counter part of the Mandelbrot
set. The two light cones are separated thus by a timelike
distance. On the real axis $\R 1_S$ things doesn't change at all.
See figure 5.

The mono quadratic case would be reobtained if one would ignore the
hyperbolic structure.

The hyperbolic case is the most interesting one, because of the
difference to the sets obtained in literature.
The asymptotic is as in the usual case. The separation results
from a deformation of the backward cone light (negative abscissa)
and a minor deformation of the forward cone. If the picture is
scaled in such a way, that the separation distance is small, one
obtains the ordinary cone structure. Points near the space like
region in the backward cone become space like during the
iteration. Points inside quadrangles which intersect the real
line are non divergent points and thus the counter part to the
mandelbrot set. The vertical cones without structure constitutes
the space like region.

Why is this dynamically interesting case called trivial?

This stems from the quadratic relation
\be
X^2-cX&=&0.
\ee
Let us assume that $c=1$\footnote{The sign does not matter in
this case.}, then one arrives at
\be
X^2&=&X,
\ee
which is a projector equation. The algebra may then be
decomposed with help of $X$ into a direct sum.
\be
1_S&=&X+(1-X)\nn \\
X(1-X)&=&0\nn \\
X^2=X && (1-X)^2=(1-X)
\ee
Therefore $X$, $(1-X)$ are pair wise annihilating primitive
idempotents.

The metric structure is then connected with
\be
[B]=[G]+[F]&=&
\left(
\begin{array}{cc}
0 & 0 \\
0 & 0
\end{array}
\right)+
\left(
\begin{array}{cc}
0   & \frac{c}{2} \\
\frac{-c}{2} & 0
\end{array}
\right),
\ee
which is a symplectic structure or equivalently\footnote{With
the isomorphism criterion from section IV.} with
\be
[B]=[G]&=&
\left(
\begin{array}{cc}
\frac{c^2}{4} & 0 \\
0              & -1
\end{array}
\right),
\ee
without any antisymmetric contribution! This corresponds to
$X^2=c^2 > 0$. The decomposition is now obtained by the projectors
\be
e_\pm&:=&\frac{1}{2c}(c\pm X)
\ee

The parameter $a$ acts as in the complex case as is seen in
figure 6. The fact that the trivial case can always be splitted
into a direct sum with diagonal multiplicative structure was
essential for the proof in \cite{Met}.

In the pure hyperbolic case ($d = 4$, $a=0$) we have for example
\be
[B]&=&
\left(
\begin{array}{cc}
1 & 0 \\
0 & -1
\end{array}
\right)
\ee
and
\be
\gamma e_1 &=& e_2\nn \\
\gamma e_2 &=& e_1,
\ee
which is not a rotation, but a space--time inversion. This
transformation flips also the orientation of $\V$, thus a
physical interpretation should be charge or parity conjugation.
But there is a continuum of such transformations.

\section*{VI Conclusion}

We showed with numerical examples, that the multi quadratic
Mandelbrot set is superior to the quadratic one. The geometric
interpretation fits in all special cases, but then with a
distinguished (pseudo) norm. The operator spinor approach is the
key step in this consideration. In a first step we considered
the two dimensional case, which has to be enlarged to higher
dimensions. Thereby the theorems from \cite{Hah} provide us that
the same structure appears as tensor factor in the
representation theory of Clifford Algebras.

The strong connection between conjugation and (pseudo) norm, as
with the geometry of the underlying vector space (module \ldots)
was shown. Thus a knowledge of the bilinear form in $\V$ provides
us with all information needed. One is able to choose even the
special ideal out of the isomorphic ones. The dependence of the
multivector structure on this choices was shown.

The field {\bf C} plays a special role, as the only one with a single
nontrivial isomorphism class. This may be the origin of the
usefulness of the complex numbers in quantum mechanics and
nonlinear classical mechanics. This fact remains in higher
dimensions.

We remarked the richness of this structure if the underlying
space is build up over rings as {\bf Z} or finite fields as {\bf
F}$_q$. An investigation in this direction should result in much
more different cases. These cases are quite interesting in
quantum theory, because they will be expected to be connected
with inequivalent representations. Normally this is achieved only
with infinitely many particles.

%\section*{Acknowledgement}
%%%%%%
% Anhang
%
\section*{Appendix}
\begin{appendix}
\setcounter{equation}{0}
\renewcommand{\theequation}{A-\arabic{equation}}
The figures are calculated with 800 $\times$ 800 points
resolution and 500 iterations in a window $[-5:5]$ for the $1_S$
(hor.) and $\gamma$ (vert.) axis. If the norms got negative, the
iteration had been stopped. The potential lines give the tendency
of reaching infinity by surmounting a given threshold in
$n$--steps. Twelve such lines are plotted. The interior of the
Mandelbrot set and the multi quadratic set consists of non
divergent points. $d=a^2+4b$ is the discriminant. $a$ is as in
the text the linear part of the quadratic relation.

\begin{figure}
\centerline{\psfig{figure=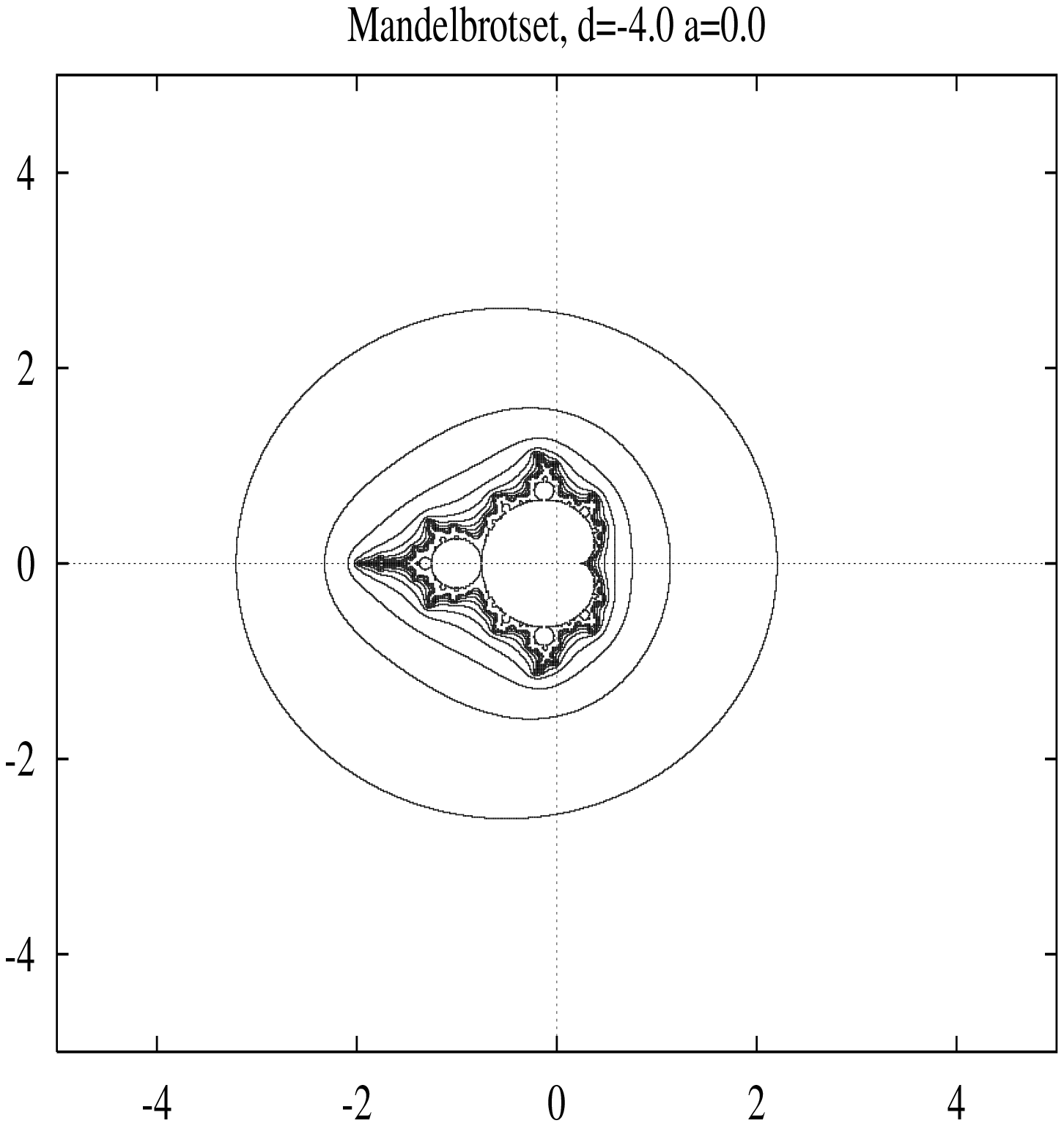,height=5.5cm,width=5.5cm}
       \hfil\psfig{figure=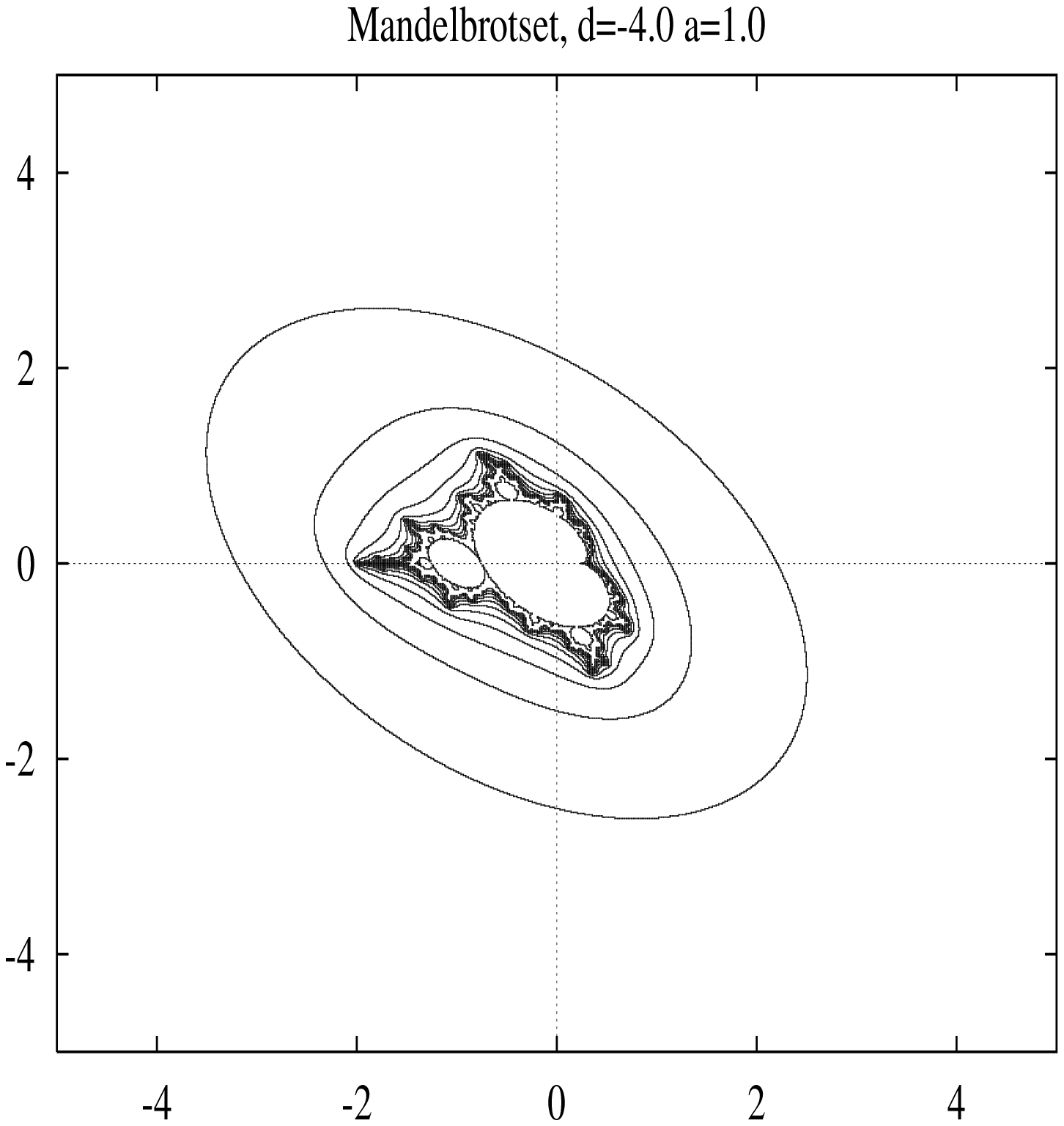,height=5.5cm,width=5.5cm}}
\centerline{\hfil Fig.1 \hfil\hfil\hfil
                  Fig.2 \hfil}
\centerline{\psfig{figure=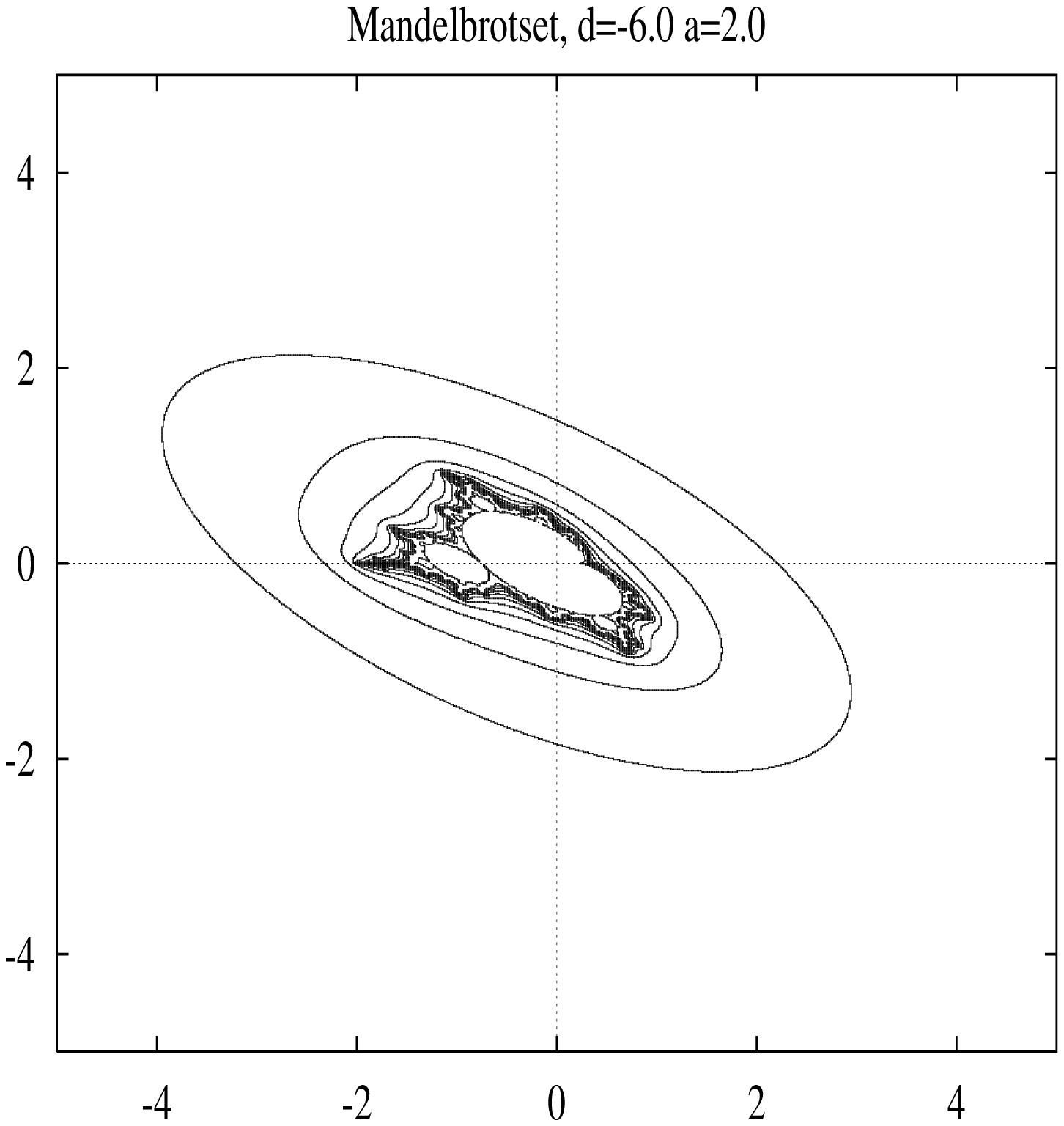,height=5.5cm,width=5.5cm}
       \hfil\psfig{figure=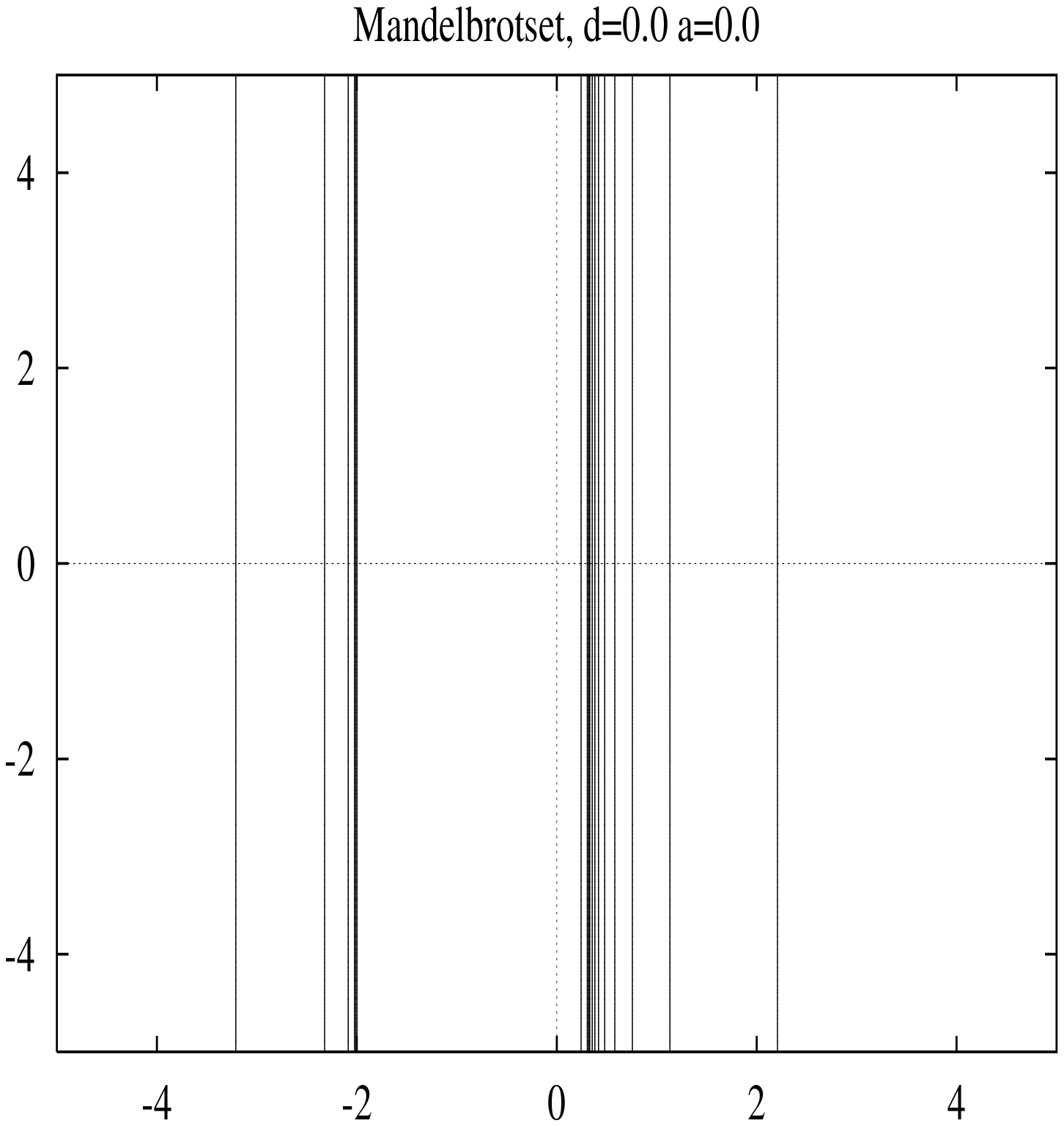,height=5.5cm,width=5.5cm}}
\centerline{\hfil Fig.3 \hfil\hfil\hfil
                  Fig.4 \hfil}
\centerline{\psfig{figure=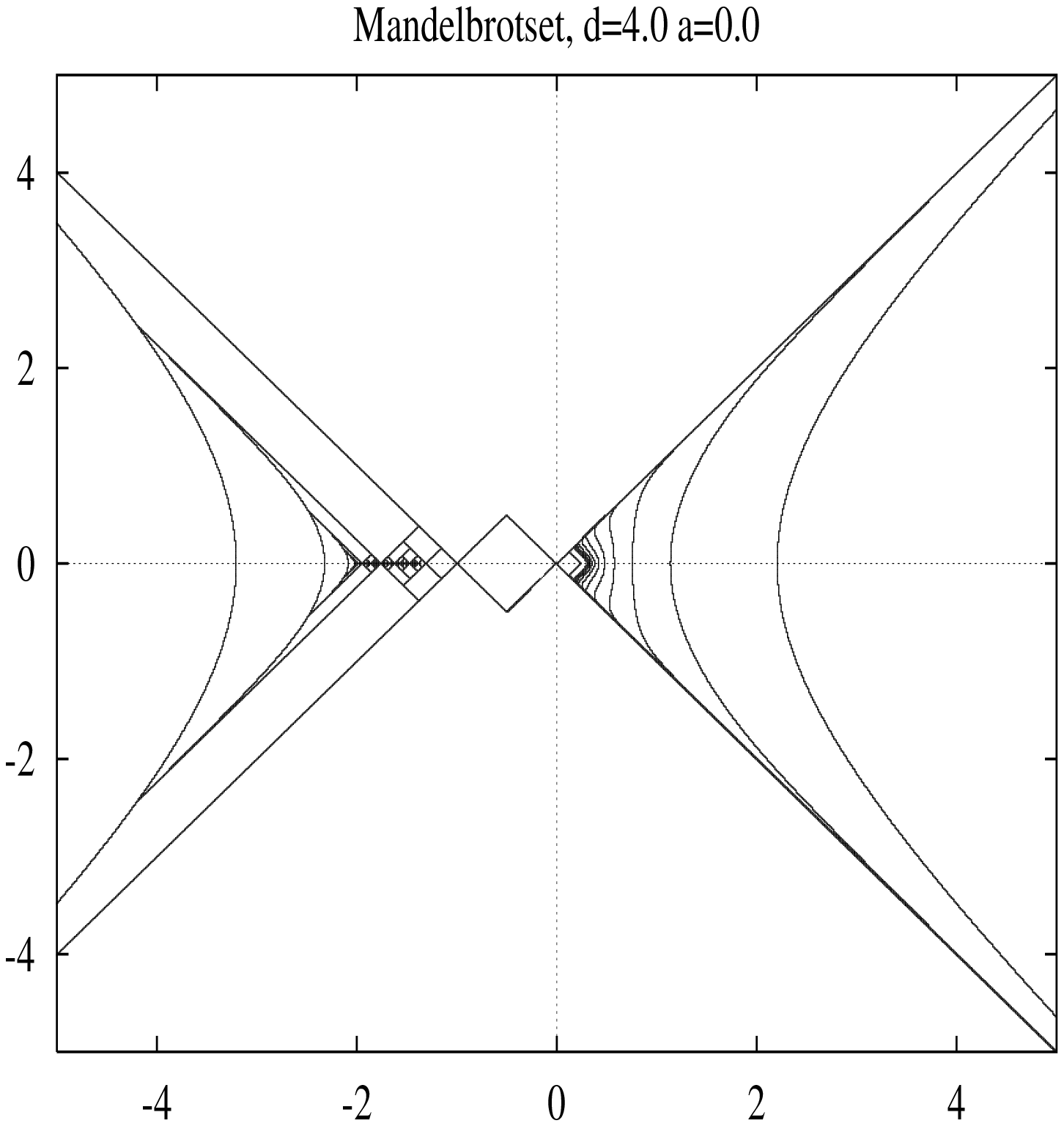,height=5.5cm,width=5.5cm}
       \hfil\psfig{figure=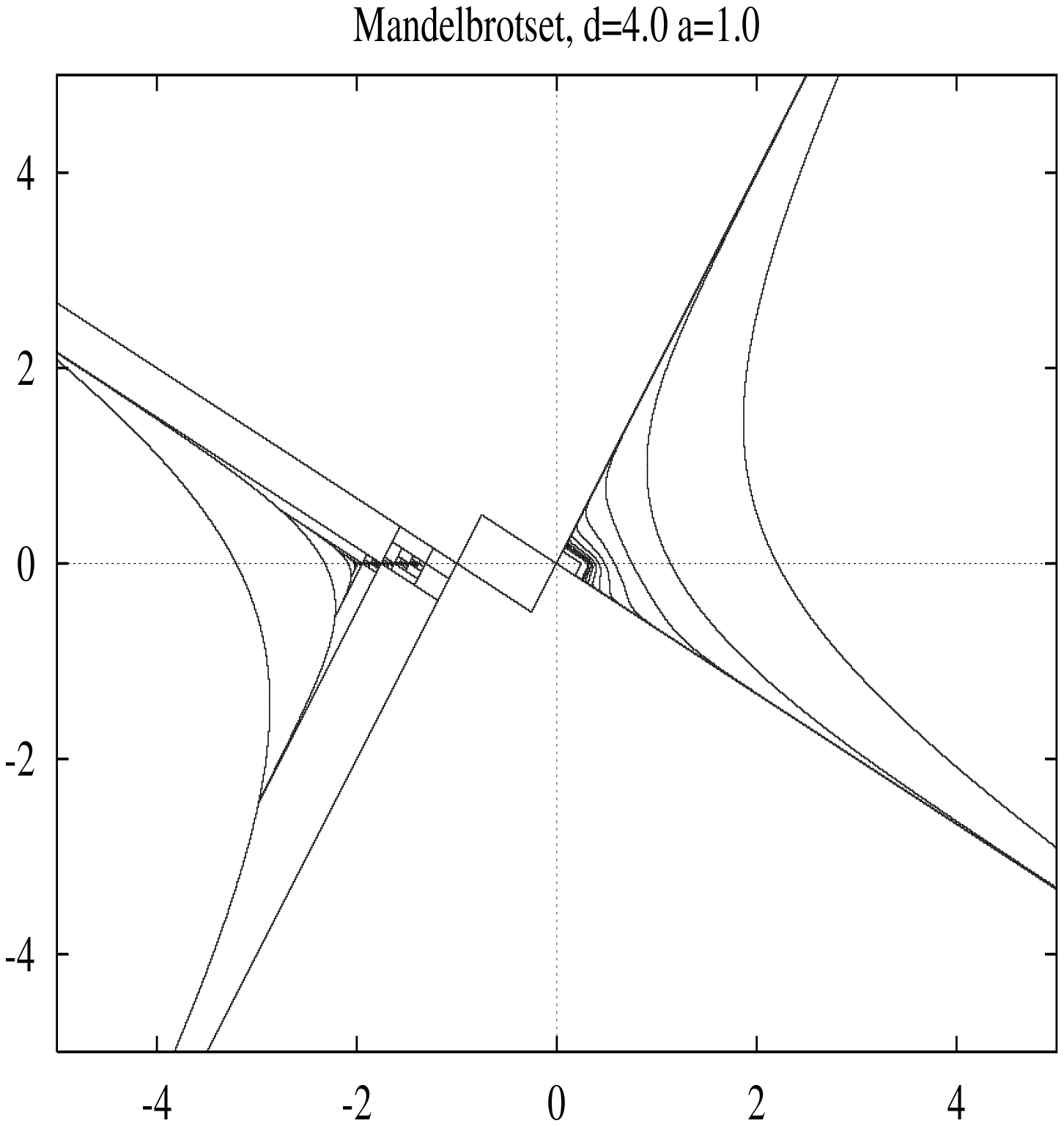,height=5.5cm,width=5.5cm}}
\centerline{\hfil Fig.5 \hfil\hfil\hfil
                  Fig.6 \hfil}
\end{figure}
\end{appendix}
%
%
% Literaturliste
%

%
%
\end{document}